\documentclass[preprint,showpacs]{revtex4}

\usepackage{epsfig}

\begin{document}

\author{Fan Wang$^{1,3}$\footnote{corresponding author:fgwang@chenwang.nju.edu.cn},
X.S.Chen$^{2,3}$, X.F. Lu$^2$, W. M. Sun$^{1,3}$, and T.Goldman$^4$}
\address{$^{1}$ Department of Physics, Nanjing University, and
Joint Center of Particle, Nuclear Physics and Cosmology, Nanjing
University and PMO, CAS, Nanjing 210093}
\address{$^{2}$ Department of Physics, Sichuan University, Chengdu, 610064}
\address{$^{3}$ Kavli Inst. for Theor. Phys. China, CAS, Beijing 100190}
\address{$^{4}$ Theoretical Division, Los Alamos National Laboratory, Los
Alamos, NM 87545, USA}

\title{The application of gauge invariance and canonical quantization to the internal
structure of gauge field systems}

\begin{abstract}

It is unavoidable to deal with the quark and gluon momentum and
angular momentum contributions to the nucleon momentum and spin in
the study of nucleon internal structure. However, we never have the
quark and gluon momentum, orbital angular momentum and gluon spin
operators which satisfy both the gauge invariance and the canonical
momentum and angular momentum commutation relation. The conflicts
between the gauge invariance and canonical quantization requirement
of these operators are discussed. A new set of quark and gluon
momentum, orbital angular momentum and spin operators, which satisfy
both the gauge invariance and canonical momentum and angular
momentum commutation relation, are proposed. The key point to
achieve such a proper decomposition is to separate the gauge field
into the pure gauge and the gauge covariant parts. The same
conflicts also exist in QED and quantum mechanics and have been
solved in the same manner. The impacts of this new decomposition to
the nucleon internal structure are discussed.
\end{abstract}

\maketitle

\section{Introduction}

In quantum physics, any observable is expressed as a Hermitian
operator in Hilbert space. The fundamental operators, such as
momentum, orbital angular momentum, spin, satisfy the canonical
momentum and angular momentum commutation relation. These
commutation relations or Lie algebras define the properties of these
operators.

Gauge invariance has been recognized as the first principle through
the development of the standard model. In classical gauge field
theory, gauge invariance principle requires any observable must be
expressed in terms of gauge invariant variable. In quantum gauge
field theory, in general one only requires the matrix elements of an
operator in between physical states to be gauge invariant. However,
one usually requires the operators themselves to be gauge invariant.
This is called the strong gauge invariance in\cite{lab1} and is
natural in QCD because all physical states are color singlet, i.e.,
$SU^c(3)$ color group gauge invariant. We will restrict our
discussion in strong gauge invariance in this paper and leave the
other possibility to the future study\cite{lab1,lab2}.

In the study of nucleon (atom) internal structure, it is unavoidable
to study the quark, gluon (electron, photon) momentum, orbital
angular momentum and spin contributions to the nucleon (atom)
momentum and spin. However, we never have the quark, gluon
(electron, photon) momentum, orbital angular momentum and spin
operators which satisfy both the gauge invariance and canonical
momentum and angular momentum commutation relation except the quark
(electron) spin. Even it has been claimed in some textbooks that one
can not define separately the photon spin and orbital angular
momentum operators \cite{lab3} and almost a proper gluon spin
operator search has been given up in the nucleon spin structure
study for the last ten years. This situation has left puzzles in
quantum mechanics, quantum electrodynamics (QED) and quantum
chromodynamics (QCD). For example, the expectation value of the
Hamiltonian of hydrogen atom is gauge dependent under a time
dependent gauge transformation\cite{lab4}, the meaning of the
multipole radiation analysis from atom to hadron spectroscopy would
be obscure if the photon spin and orbital angular momentum operators
were not well defined, especially the parity of these microscopic
states determined from the multipole radiation analysis would be
obscure, there will be no way to compare the measured gluon spin
contribution to nucleon spin with the theoretically calculated one
if one has not a proper gluon spin operator, etc.

 In section II the conflict between gauge invariance and canonical
quantization of the usual quark gluon (electron photon) momentum,
orbital angular momentum and spin operators are discussed from the
simple quantum mechanics of a charged particle moving in an
electromagnetic field to those of quark and gluon in QCD. In the
third section a new set of momentum, orbital angular momentum and
spin operators, which satisfy both the gauge invariance and
canonical momentum and angular momentum commutation relation, are
given. The key point to achieve this is to separate the gauge field
into pure gauge and gauge covariant (invariant) parts. The possible
impacts of these modification to the nucleon internal structure will
be discussed in section IV. The last section is a summary and a
prospect of further studies.

\section{Gauge invariance and canonical quantization
of the momentum and angular momentum operators of the Fermion and
gauge field parts}

The conflict in the application of gauge invariance and canonical
quantization to the momentum and orbital angular momentum operators
of a charged particle moving in the electromagnetic field, a U(1)
Abelian gauge field, has existed in quantum mechanics since the
establishment of gauge invariance principle. Starting from the
Lagrangian of a non-relativistic charged particle with mass m,
velocity $\vec{v}$ and charge e moving in an electromagnetic field
$A^{\mu}=(A^0, \vec{A})$,
\begin{eqnarray}
\mathcal{L}(m,\vec{v},e,A_{\mu}) &=& \frac{1}{2m}
(m\vec{v})^2-e(A^0-\vec{v}\cdot\vec{A}),
\end{eqnarray}
one obtains the canonical momentum, orbital angular momentum and
Hamiltonian
\begin{equation}
\vec{p}=m\vec{v}+e\vec{A},~~~ \vec{L}=\vec{r}\times\vec{p},~~~H=
\frac{1}{2m} (\vec{p}-e\vec{A})^2 + eA^0.
\end{equation}
All of these three classical dynamical variables are gauge dependent
and so are not observables in classical gauge theory. In the
coordinate representation, the momentum $\vec{p}$ is quantized as
\begin{equation}
\vec{p}= \frac{\vec{\nabla}}{i}.
\end{equation}
no matter what kind of gauge is fixed on even though the classical
canonical momentum is gauge dependent. The orbital angular momentum
and Hamiltonian are quantized by replacing $\vec{p}$ by
$\frac{\vec{\nabla}}{i}$. These quantized momentum and angular
momentum operators satisfy the canonical commutation relation or the
Lie algebra:
\begin{equation}
[p_l,p_m]=0,~~~[L_l,L_m]=i\epsilon_{lmn}L_n,~~~[p_l,L_m]=
i\epsilon_{lmn}p_n,~~~ l,m,n=1,2,3,
\end{equation}
where $\epsilon_{lmn}$ is the rank-three totally antisymmetric
tensor and $\epsilon_{123}=1$. In general, the $[p_l,H]\neq 0$,
which is different from the Poincar$\acute{e}$ algebra of the total
momentum $P_l, (l=1,2,3)$ and total H of the whole system where
$[P_l,H]=0$.

However, after a gauge transformation,
\begin{equation}
\psi'=e^{-ie\omega(x)}\psi,~~~A'^{\mu}=A^{\mu}+\partial^{\mu}\omega(x),
\end{equation}
the matrix elements of the above operators transform as follows,
\begin{eqnarray}
\langle\psi'|\vec{p}|\psi'\rangle &=&
\langle\psi|\vec{p}|\psi\rangle
-e\langle\psi|\vec{\nabla}\omega(x)|\psi\rangle,\nonumber\\
\langle\psi'|\vec{L}|\psi'\rangle &=&
\langle\psi|\vec{L}|\psi\rangle
-e\langle\psi|\vec{r}\times\vec{\nabla}\omega(x)|\psi\rangle,\nonumber\\
\langle\psi'|H'|\psi'\rangle &=&
\langle\psi|H|\psi\rangle+e\langle\psi|\partial_t\omega(x)|\psi\rangle.
\end{eqnarray}
It is obvious that the matrix elements of these three operators are
all gauge dependent. Therefore they are not measurable and so these
operators do not correspond to observables. This problem has left in
quantum mechanics since the gauge principle was proposed.

The relativistic version of quantum mechanics has the same problem.
The gauge dependence of the expectation value of the Hamiltonian of
a charged particle moving in an electromagnetic field under a time
dependent gauge transformation was discussed by T.
Goldman\cite{lab4}.

This conflict is carried over to QED. Starting from a QED
Lagrangian,
\begin{eqnarray}
\mathcal{L} &=&
\bar{\psi}[i\gamma^{\mu}(\partial_{\mu}+ieA_{\mu})-m]\psi
-\frac{1}{4}F_{\mu\nu}F^{\mu\nu}, \nonumber\\
F_{\mu\nu} &=& \partial_{\mu}A_{\nu}-\partial_{\nu}A_{\mu}.
\end{eqnarray}
By means of the Noether theorem one obtains the momentum and angular
momentum operators as follows:
\begin{equation}
\vec{P}=\vec{P}_e+\vec{P}_{ph}=\int
d^3x\psi^{\dag}\frac{\vec{\nabla}}{i}\psi+\int d^3x
E^{i}\vec{\nabla}A^{i},\label{PQED}
\end{equation}
\begin{eqnarray}
\vec{J} &=&
\vec{S}_e+\vec{L}_e+\vec{S}_{ph}+\vec{L}_{ph} \nonumber\\
&& =\int d^3x\psi^{\dag}\frac{\vec{\Sigma}}{2}\psi+\int
d^3x\vec{x}\times\psi^{\dag}\frac{\vec{\nabla}}{i}\psi +\int
d^3x\vec{E}\times\vec{A}+\int d^3x\vec{x}\times E^i\vec{\nabla}A^i.
\label{JQED}
\end{eqnarray}

Here $\Sigma^j=\frac{i}{2}\epsilon_{jkl}\gamma^k\gamma^l$. These
electron and photon momentums, orbital angular momentums and spins,
after quantization, satisfy momentum and angular momentum Lie
algebra. However, they are not gauge invariant except the electron
spin.

The multipole radiation analysis is the basis of atomic, molecular,
nuclear and hadron spectroscopy. The multipole field is based on the
decomposition of the electromagnetic field into field with definite
orbital angular momentum and spin quantum numbers. If the photon
spin and orbital angular momentum operators were gauge dependent,
then the physical meaning of the multipole field would be obscure,
especially the parity of these microscopic states determined by the
measurement of the orbital angular momentum quantum number of the
multipole radiation field would be obscure.

QCD has the same problem as QED. The quark gluon momentum, orbital
angular momentum and spin operators derived from QCD Lagrangian by
Noether theorem have the same form as those of electron and photon,
Eqs.(\ref{PQED},\ref{JQED}), if one omits the color indices. They
satisfy the momentum and angular momentum Lie algebra but they are
not gauge invariant except the quark spin.

Because of the lack of gauge invariant quark, gluon momentum
operators, the present operator product expansion (OPE) used the
following two operators as quark and gluon momentum operators (the
color indices are suppressed),
\begin{eqnarray}
\vec{P} &=& \vec{\mathcal{P}}_q+\vec{\mathcal{P}}_g =\int
d^3x\psi^{\dag}\frac{\vec{D}}{i}\psi+\int d^3x
\vec{E}\times\vec{B}, \nonumber\\
\vec{D} &=& \vec{\nabla}-ig\vec{A}.
\end{eqnarray}
Both the quark and gluon "momentum" operators $\vec{\mathcal{P}}_q$
and $\vec{\mathcal{P}}_g$ defined in Eq.(10) are gauge invariant but
neither the quark "momentum" $\vec{\mathcal{P}}_q$ nor the gluon
"momentum" $\vec{\mathcal{P}}_g$ satisfies the momentum algebra, for
example,
\begin{equation}
[D_l,D_m]=-ig(\partial_l A_m-\partial_m
A_l)-ig^2C_{abc}A^a_lA^b_mT^c,
\end{equation}
where $C_{abc}$ is the $SU(3)$ group structure constant. The
$\vec{\mathcal{P}}_g$ does not satisfy the momentum algebra either
in the interacting quark-gluon field, i.e., QCD case. Therefore
neither the $\vec{\mathcal{P}}_q$ nor the $\vec{\mathcal{P}}_g$ used
in the OPE is the real momentum operator.

A gauge invariant decomposition of quark, gluon angular momentum has
been proposed\cite{lab5} and used in the study of nucleon spin
structure in the last ten years,
\begin{eqnarray}
\vec{J} &=& \vec{S}_q+\mathcal{\vec{L}}_q+\mathcal{\vec{J}}_g  \nonumber\\
&& =\int d^3x\psi^{\dag}\frac{\vec{\Sigma}}{2}\psi+\int d^3x
\vec{x}\times\psi^{\dag}\frac{\vec{D}}{i}\psi+\int
d^3x\vec{x}\times(\vec{E}\times\vec{B}),\label{JQCD}
\end{eqnarray}
here the color indices have been suppressed. Each term in this
decomposition is gauge invariant, however, they do not satisfy the
angular momentum algebra except the quark spin. In addition, there
is no gluon spin in this decomposition. Experimentally the gluon
spin contribution to nucleon spin is under intensive study, PHENIX,
STAR, COMPASS, HERMES, and others are measuring the gluon spin
contribution to nucleon spin. Neither the decomposition,
Eq.(\ref{JQED}), nor the decomposition, Eq.(\ref{JQCD}), gives a
gluon spin operator which satisfies both gauge invariance and
angular momentum algebra. There is also no quark, gluon orbital
angular momentum operator which satisfies both gauge invariance and
orbital angular momentum algebra. These situations hindered the
study of the nucleon spin structure.

\section{A new set of momentum, orbital angular momentum and spin
operators for the Fermion and gauge field parts}

\subsection{Decomposing the gauge field $A_{\mu}$ into pure gauge part
${A}_{pure}$ and gauge invariant (covariant) part ${A}_{phys}$}

Let us start from the simpler QED case. It is well known that to use
gauge potential $A_{\mu}$ to describe the electromagnetic field the
$A_{\mu}$ is not unique, i.e., there is gauge freedom. Under a gauge
transformation,
\begin{equation}
A'^{\mu}=A^{\mu}+\partial^{\mu}\omega(x),
\end{equation}
one obtains a new gauge potential $A'^{\mu}$ from $A^{\mu}$.
$A^{\mu}$ and $A'^{\mu}$ describe the same electromagnetic field,
\begin{equation}
F_{\mu\nu}=\partial_{\mu}A_{\nu}-\partial_{\nu}A_{\mu}=\partial_{\mu}A'_{\nu}-\partial_{\nu}A'_{\mu}.
\end{equation}
Such a gauge freedom is necessary because the gauge potential
$A^{\mu}$ plays two role in gauge field theory: the first is to
provide a pure gauge field $A_{pure}$ to compensate the induced
field due to the phase change in a local gauge transformation of the
Fermion field $\psi'(x)=e^{-ie\omega(x)}\psi(x)$ which must be
varied with the arbitrarily changed phase parameter $\omega(x)$; the
second is to provide a physical field $A_{phys}$ for the physical
interaction between Fermion field and gauge field which should be
gauge invariant (covariant) under gauge transformation. The pure
gauge potential $A_{pure}$ should not contribute to electromagnetic
field,
\begin{equation}
F^{\mu\nu}_{pure}=\partial^{\mu}A^{\nu}_{pure}-\partial^{\nu}A^{\mu}_{pure}=0.
\end{equation}
This equation can not fix the $A_{pure}$ uniquely. One has to find
additional condition to fix it. The spatial part of Eq.(15) is
\begin{equation}
\nabla\times \vec{A}_{pure}=0,
\end{equation}
which means $\vec{A}_{pure}$ does not contribute to magnetic field.
This equation can be expressed in another form,
\begin{equation}
\nabla\times\vec{A}_{phys}=\nabla\times\vec{A}.
\end{equation}
A natural choice of the additional condition in QED case is
\begin{equation}
\nabla\cdot\vec{A}_{phys}=0,
\end{equation}
which is the transverse wave condition and we know that the ${\vec
A}_{phys}$ part is the physical one from the Coulomb gauge
quantization. Combining these two conditions, Eqs.(17) and (18),
under the natural boundary condition,
\begin{equation}
\vec{A}_{phys}(|x|\rightarrow\infty)=0,
\end{equation}
for any given set of gauge field $\vec{A}$, one can decompose it
uniquely as follows,
\begin{equation}
\vec{A}=\vec{A}_{pure}+\vec{A}_{phys},
\end{equation}
where
\begin{eqnarray}
\vec{A}_{phys}(x) &=& \vec{\nabla}\times\frac{1}{4\pi}\int
d^3x'\frac{\vec{\nabla}'\times\vec{A}(x')}{|\vec{x}-\vec{x'}|},\nonumber\\
\vec{A}_{pure}(x) &=& \vec{A}-\vec{A}_{phys}(x).
\end{eqnarray}
We have to emphasize that for fixed $\vec{A}(x)$, the integration
can be done and the obtained $\vec{A}_{phys}(x)$ is a local function
of space-time x and a functional of $A_{\mu}(x)$. It is easy to
prove that these two parts transform as follows in a gauge
transformation Eq.(13),
\begin{eqnarray}
\vec{A}'_{phy} & = & \vec{A}_{phy}, \nonumber \\
\vec{A}'_{pure} & = & \vec{A}_{pure}-\vec{\nabla}\omega(x).
\end{eqnarray}
The time component $A^0$ can be decomposed in the same manner. From
the condition $F^{i0}_{pure}=0$, one obtains
\begin{eqnarray}
\partial_i{A}^0_{phys} &=&
\partial_i{A}^0+\partial_t(A^i-A^i_{phys}), \nonumber\\
A^0_{phys} &=& \int_{-\infty}^x
dx^i(\partial_i{A}^0+\partial_t{A}^i-\partial_t{A}^i_{phys}).
\end{eqnarray}
The $A^{\mu}_{pure}=A^{\mu}-A^{\mu}_{phys}$ can also be obtained
from Eq.(16), (18) and (23) directly,
\begin{eqnarray}
\vec{A}_{pure} &=& \vec{\nabla}\phi(x),\nonumber\\
\phi(x) &=& -\frac{1}{4\pi}\int
d^3x'\frac{{\nabla}'\cdot\vec{A}(x')}{|\vec{x}-\vec{x'}|}+\phi_0(x),\nonumber\\
A^0_{pure} &=& \partial_t\phi(x),
\end{eqnarray}
where $\phi_0(x)$ satisfies the condition,
\begin{equation}
{\nabla}^2\phi_0(x)=0,
\end{equation}
and is determined by the boundary condition.

To decompose the gauge potential $A_{\mu}=A^a_{\mu}T^a$ for the
gluon field is more complicated than QED case. We first define the
pure gauge potential $A^{\mu}_{pure}$ (hereafter we omit the color
indices if not necessary) by the same condition, i.e., it does not
contribute to color electromagnetic field,
\begin{equation}
F^{\mu\nu}_{pure}=\partial^{\mu}A^{\nu}_{pure}-\partial^{\nu}A^{\mu}_{pure}
+ig[A^{\mu}_{pure},A^{\nu}_{pure}]=0.
\end{equation}
In order to make this defining condition looks similar to Eq.(16),
we introduce a notation,
\begin{eqnarray}
\vec{D}_{pure} &=& \vec{\nabla}-ig\vec{A}_{pure}, \nonumber\\
\vec{D}_{pure}\times\vec{A}_{pure} &=&
\vec{\nabla}\times\vec{A}_{pure}-ig\vec{A}_{pure}\times\vec{A}_{pure}=0.
\end{eqnarray}
The additional condition is even more complicated, i.e., one does
not have a natural choice as Eq.(18) in QED. We make the following
choice\cite{lab6},
\begin{eqnarray}
\vec{\mathcal{D}}_{pure} &=&
\vec{\nabla}-ig[\vec{A}_{pure},~~]\nonumber\\
\vec{\mathcal{D}}_{pure}\cdot\vec{A}_{phys} &=&
\vec{\nabla}\cdot\vec{A}_{phys}-ig[{A}^i_{pure},{A}^i_{phys}]=0.
\end{eqnarray}
The summation over the vector component i has been assumed in the
above equation and the following ones. Please note that in the above
adjoint representation of the new covariant derivative operator
$\vec{\mathcal{D}}$, the bracket $[A^i_{pure},A^i_{phys}]$ is not
the quantum bracket but a color $SU^c(3)$ group one,
\begin{equation}
[A^i_{pure},A^i_{phys}]=iC_{abc}A^{ib}_{pure}A^{ic}_{phys}T^a.\nonumber
\end{equation}
These equations can be rewritten as follows,
\begin{eqnarray}
\vec{\nabla}\cdot\vec{A}_{phys} &=&
ig[{A}^i-{A}^i_{phys},{A}^i_{phys}]=ig[{A}^i,{A}^i_{phys}],\nonumber\\
\vec{\nabla}\times\vec{A}_{phys} &=&
\vec{\nabla}\times\vec{A}-ig(\vec{A}-\vec{A}_{phys})\times(\vec{A}-\vec{A}_{phys}),\\
\partial_i{A}^0_{phys} &=&
\partial_i{A}^0+\partial_t(A^i-A^i_{phys})-ig[A^i-A^i_{phys},A^0-A^0_{phys}].\nonumber
\end{eqnarray}
These equations can be solved perturbatively: in the zeroth order,
i.e., assuming $g=0$, these equations are the same as those of QED,
one can obtain the zeroth order solution; then taking into account
the nonlinear coupling through iteration one obtains a perturbative
solution as a power expansion in g.

If one assumes a trivial boundary condition for the pure gauge field
$A_{pure}$, then one can use the following equations to obtain a
perturbative solution too,
\begin{eqnarray}
\vec{\nabla}\times\vec{A}_{pure} &=&
ig\vec{A}_{pure}\times\vec{A}_{pure}, \nonumber\\
\vec{\nabla}\cdot\vec{A}_{pure} &=&
\vec{\nabla}\cdot\vec{A}-ig[A^i_{pure},A^i],\\
\partial_iA^0_{pure} &=& -\partial_t
A^i_{pure}+ig[A^i_{pure},A^0_{pure}]. \nonumber
\end{eqnarray}
Under a gauge transformation,
\begin{eqnarray}
\psi' &=& U\psi, \nonumber\\
A'_{\mu} &=& UA_{\mu}U^{\dag}-\frac{i}{g}U\partial_{\mu}U^{\dag},
\end{eqnarray}
where $U=e^{-ig\omega}$. The $\vec{A}_{pure}$ and $\vec{A}_{phys}$
will be transformed as
\begin{eqnarray}
\vec{A'}_{phys} &=& U\vec{A}_{phys}U^{\dag}, \nonumber\\
\vec{A'}_{pure} &=&
U\vec{A}_{pure}U^{\dag}-\frac{i}{g}U\partial_{\mu}U^{\dag}.
\end{eqnarray}

\subsection{Quantum mechanics}

We have mentioned in the introduction part that even in quantum
mechanics, there are already puzzles related to the fundamental
operators, the matrix elements of canonical momentum, orbital
angular momentum and Hamiltonian of a charged particle moving in an
electromagnetic field are all not gauge invariant. In order to get
rid of these puzzles, gauge invariant operators have been
introduced,
\begin{equation}
\mathcal{\vec{P}} =
\vec{p}-e\vec{A},~~~\mathcal{\vec{L}}=\vec{x}\times\mathcal{\vec{P}}.
\end{equation}
It is easy to check that the matrix elements of these operators are
gauge invariant. However, as we have pointed out in Eq.(11), that
the gauge invariant "momentum" $\mathcal{\vec{P}}$ does not satisfy
the canonical momentum Lie algebra, so they are not the proper
momentum. The gauge invariant "orbital angular momentum"
$\mathcal{\vec{L}}$ does not satisfy the angular momentum Lie
algebra either.

Based on our proposed gauge field decomposition in the above
section, we introduce another set of momentum and orbital angular
momentum operators which satisfy both gauge invariance and the
corresponding commutation relation,
\begin{eqnarray}
\vec{p}_{pure} &=& \vec{p}-e\vec{A}_{pure}, \nonumber\\
\vec{L}_{pure} &=& \vec{x}\times\vec{p}_{pure}.
\end{eqnarray}

The long-standing puzzle, the gauge non-invariance of the
expectation value of the Hamiltonian\cite{lab4} can be solved in the
same manner. For the non-relativistic quantum mechanics, the new
Hamiltonian is
\begin{equation}
H=\frac{(\vec{p}-e\vec{A}_{pure}-e\vec{A}_{phys})^2}{2m}+e(A^0-\partial_t\phi(x)).
\end{equation}
The last term is a pure gauge term, it cancels the unphysical energy
appearing in $eA^0$ induced by the pure gauge term and then
guarantees the expectation value of this Hamiltonian gauge
invariant. It is a direct extension of Eq.(35) to the zeroth
momentum component.

The Dirac Hamiltonian has the same unphysical energy part and has to
be canceled in the same manner as that for the Schr$\ddot{o}$dinger
Hamiltonian. Here we have done a check: starting from a QED
Lagrangian with both electron and proton, under the infinite proton
mass approximation, we derived the Dirac equation of electron and
the gauge invariant Hamiltonian of the electron part and verified
the difference between the Dirac Hamiltonian obtained from the Dirac
equation and the gauge invariant one from the energy-momentum
tensor.

Our study shows that the canonical momentum, orbital angular
momentum and the Hamiltonian used in quantum mechanics are not
observables, one must subtract the pure gauge part, the unphysical
one, from these operators as we did in Eq.(35) and (36) to obtain
the observable ones. In Coulomb gauge, where the $A^{\mu}_{pure}=0$,
the momentum, orbital angular momentum and Hamiltonian operators in
Eqs.(35,36) are simplified to the familiar form used in quantum
mechanics. This justifies the quantum mechanics calculation with
Coulomb gauge.

\subsection{QED}

We have explained that the momentum and angular momentum operators
of the Fermion and gauge field part, Eqs.(8) and (9), derived from
the QED Lagrangian by means of Noether theorem are not gauge
invariant except the Fermion spin. One can obtain a gauge invariant
decomposition by adding a surface term to Eqs.(8) and (9) or from
the Belinfante symmetric energy-momentum tensor,
\begin{eqnarray}
\vec{P} &=& \vec{\mathcal{P}_e}+\vec{\mathcal{P}_{ph}} = \int
d^3x\psi^{\dag}\frac{\vec{D}}{i}\psi+\int d^3x\vec{E}\times\vec{B}\\
\vec{J} &=&
\vec{S}_e+\mathcal{\vec{L}}_e+\mathcal{\vec{J}}_{ph}=\int
d^3x\psi^{\dag}\frac{\vec{\Sigma}}{2}\psi+\int d^3x
\vec{x}\times\psi^{\dag}\frac{\vec{D}}{i}\psi+\int
d^3x\vec{x}\times(\vec{E}\times\vec{B}).
\end{eqnarray}
There are two problems with this decomposition: (1),
$\mathcal{\vec{P}}_e$, $\mathcal{\vec{P}}_{ph}$,
$\mathcal{\vec{L}}_e$ and $\mathcal{\vec{J}}_{ph}$ do not satisfy
the momentum and angular momentum commutation relations even though
in free electromagnic field theory the photon momentum
$\mathcal{\vec{P}}_{ph}$ and angular momentum
$\mathcal{\vec{J}}_{ph}$ do; (2), there is no separate photon spin
and orbital angular momentum operators and this feature will ruin
the multipole radiation analysis as we discussed in the second
section.

Based on the decomposition of the gauge potential into pure gauge
and the physical parts, Eq.(20), we obtain the following
decomposition,
\begin{eqnarray}
\vec{P} &=& \vec{P}^{p}_e+\vec{P}^{p}_{ph},\\
&& =\int d^3x\psi^{\dag}\frac{\vec{D}_{pure}}{i}\psi+\int d^3x
E^i\vec{\mathcal{D}}_{pure}A^i_{phys}. \nonumber
\end{eqnarray}
\begin{eqnarray}
\vec{J} &=& \vec{S}_e+\vec{L}^{p}_e+\vec{S}^{p}_{ph}+\vec{L}^{p}_{ph} \\
&& =\int d^3x\psi^{\dag}\frac{\vec{\Sigma}}{2}\psi+\int
d^3x\vec{x}\times\psi^{\dag}\frac{\vec{D}_{pure}}{i}\psi
 +\int d^3x \vec{E}\times\vec{A}_{phys}+\int d^3x \vec{x}\times
E^i\vec{\mathcal{D}}_{pure}A^i_{phys}. \nonumber
\end{eqnarray}

Here the operator $\vec{D}_{pure}$ and $\vec{\mathcal{D}}_{pure}$
are the same as given in Eqs.(27) and (28) but with g replaced by e.
Because of the Abelian property of the $U(1)$ gauge field, the
adjoint representation of the operator $\vec{\mathcal{D}}$ is
simplified to be a simple $\vec{\nabla}$. It is not hard to check
that each operator in the above decomposition, Eq.(39) and (40) is
gauge invariant and satisfies the momentum, angular momentum
commutation relation. In Coulomb gauge, the operators in Eqs.(39,40)
are simplified to operator forms in Eqs.(8,9). This justifies that
the gauge non-invariant operators of Eqs.(8,9) can be used in
Coulomb gauge to get the gauge invariant results.

The photon spin and orbital angular momentum operators are well
defined as shown in Eq.(40). The multipole radiation analysis is
theoretically sound now as it should be.

\subsection{QCD}

One can copy results for QED, the Eqs.(39,40), to QCD to obtain the
quark, gluon momentum, orbital angular momentum and spin operators
which satisfy both the gauge invariance and the canonical momentum
and angular momentum commutation relations.

A decomposition of the form of Eq.(10,12) has been used in the
nucleon spin structure study for the last ten years\cite{lab5}. Each
operator in those decomposition is gauge invariant and so
corresponds to observable. However, because they do not satisfy the
momentum, angular momentum Lie algebra so the measured ones are not
the quark, gluon momentum and orbital angular momentum and can not
be connected to those used in hadron spectroscopy.

The gluon spin operator had been searched for more than ten years in
the nucleon spin structure study and no satisfactory one was
obtained. Now one has the gluon spin operator $\vec{S}^{p}_g=\int
d^3x\vec{E}\times\vec{A}_{phys}$ which can be used to calculate the
matrix element for a polarized nucleon state $|N(p,s)\rangle$ to
obtain the gluon spin contribution to nucleon spin and compare it
with the measured ones.

\section{Reexamination of the nucleon internal structure}

For the past years the nucleon internal structure has been studied
based on momentum, angular momentum operators given in Eq.(10,12).
These operators are gauge invariant but do not satisfy the momentum
and angular momentum Lie algebra. This led to a distorted picture of
the nucleon internal structure. For example, that the quark and
gluon carry half of the nucleon momentum in the asymptotic limit has
been a deeply rooted picture of nucleon internal momentum structure.
Using the new quark, gluon momentum operator we recalculated their
scale evolution and obtained the new mixing matrix\cite{lab7},
\begin{eqnarray}
\gamma^P&=&-\frac{\alpha_s}{4\pi}\left(
\begin{array}{cc}-\frac{2}{9}n_g&\frac{4}{3}n_f\\
\frac{2}{9}n_g&-\frac{4}{3}n_f\end{array} \right),
\end{eqnarray}
which gives the new asymptotic limit for the renormalized gluon
momentum,
\begin{equation}
\vec{P}^R_g=\frac{\frac{1}{2}n_g}{\frac{1}{2}n_g+3n_f}\vec{P}_{total}.
\end{equation}
For typical gluon number $n_g=8$ and quark flavor number $n_f=5$,
the above equation gives
$\vec{P}^R_g\simeq\frac{1}{5}\vec{P}_{total}$. This is distinctly
different from the renowned results
$\vec{P}^R_g\simeq\frac{1}{2}\vec{P}_{total}$. This latter result is
obtained from the mixing matrix,
\begin{eqnarray}
\gamma^P&=&-\frac{\alpha_s}{4\pi}\left(
\begin{array}{cc}-\frac{8}{9}n_g&\frac{4}{3}n_f\\
\frac{8}{9}n_g&-\frac{4}{3}n_f\end{array} \right),
\end{eqnarray}
which is obtained by means of the quark and gluon momentum operators
given in Eq.(10). The mixing matrix element of Eq.(43) leads to the
well known asymptotic limit,
\begin{equation}
\vec{\mathcal{P}}^R_g=\frac{2n_g}{2n_g+3n_f}\vec{P}_{total}.
\end{equation}
However, the $\vec{\mathcal{P}_g}$ and $\vec{\mathcal{P}_q}$ used in
this quark gluon momentum scale evolution calculation are not the
proper momentum operators, part of the quark momentum had been
shifted to the gluon and gave the superficial large gluon momentum
contribution to nucleon momentum.

The asymptotic nucleon spin structure\cite{lab8} is obtained based
on the decomposition Eq.(9), a QED analog of QCD angular momentum
decomposition. The authors had pointed out that the quark and gluon
orbital angular momentum and the gluon spin operators are not gauge
invariant.
As we have mentioned in the beginning, in the present gauge field
theory an observable must be expressed in terms of a gauge invariant
operator. The gauge dependent operators used in this
analysis\cite{lab8} are not measurable ones. Therefore this
asymptotic limit of nucleon spin content should be reexamined.

Another nucleon internal structure parameters are the parton
distribution function (PDF). For example, the quark PDF in a target
A is defined as,
\begin{eqnarray}
\mathcal{P}_{q/A}(\xi)=\frac 12 \int_{-\infty}^{\infty}
\frac{dx^-}{2\pi} e^{-i\xi P^+x^-}\langle \bar\psi(0,x^-,0_\bot)
\gamma^+\mathcal{P}\exp\{ig\int_0^{x^-}dy^- A^+(0,y^-,0_\bot)\}
\psi(0)\rangle_A, \label{pdfq}
\end{eqnarray}
where a gauge link (Wilson line) is inserted to achieve the gauge
invariance. Based on our gauge field decomposition discussed in the
third section, the above gauge link not only includes the necessary
pure gauge $A_{pure}$ part to achieve the gauge invariance, but also
includes the physical part $A_{phys}$ which induced a physical
coupling and makes the PDF defined in Eq.(45) an
interaction-involving one. The interaction term is more clear in the
momentum relation,
\begin{equation}
\int_{-\infty}^{\infty} d\xi \xi \mathcal
{P}_{q/A}(\xi)=\frac{1}{2(P^+)^2}\langle \bar \psi \gamma^+i
D^+\psi\rangle _A .
\end{equation}
This is just the + component of $\vec{\mathcal{P}}_q$ in Eq.(10).
Here the gauge field in $D^+$ originates exactly from the gauge link
in Eq.(45).

To obtain a gauge invariant quark PDF, a gauge link with the pure
gauge part is enough,
\begin{eqnarray}
P_{q/A}(\xi)= \frac 12 \int_{-\infty}^{\infty} \frac{dx^-}{2\pi}
e^{-i\xi P^+x^-}\langle P|\bar\psi(0,x^-,0_\bot) \gamma^+
\mathcal{P}\exp\{ig\int_0^{x^-}dy^- A_{\rm pure}^+(0,y^-,0_\bot)\}
\psi(0)|P \rangle _A,
\end{eqnarray}
this PDF will not include the redundant physical gauge interaction
and the integration gives the real quark momentum defined in Eq.(39)
(the QCD quark and gluon momentums have exactly the same expression
as those of QED, only the subscript e and ph are replaced by q and
g).
\begin{equation}
\int_{-\infty}^{\infty} d\xi \xi P
_{q/A}(\xi)=\frac{1}{2(P^+)^2}\langle \bar \psi \gamma^+ i D^+_{\rm
pure}\psi\rangle _A .
\end{equation}
Analogously, the conventional gluon PDF
\begin{eqnarray}
\mathcal{P}_{g/A}(\xi)= \frac{1}{\xi P^+}\int_{-\infty}^{\infty}
\frac{dx^-}{2\pi} e^{-i\xi P^+x^-}\langle F^{+\nu}(0,x^-,0_\bot)
\mathcal{P}\exp\{ig\int_0^{x^-}dy^- A^+(0,y^-,0_\bot)\} F_\nu
^{~+}(0)\rangle _A,
\end{eqnarray}
can be replaced accordingly as
\begin{eqnarray}
P_{g/A}(\xi)= \int_{-\infty}^{\infty}  \frac{dx^-}{2\pi} e^{-i\xi
P^+x^-}\langle F^{+i}(0,x^-,0_\bot)
\mathcal{P}\exp\{ig\int_0^{x^-}dy^- A_{\rm pure}^+(0,y^-,0_\bot)\}
A^i_{\rm phys}(0) \rangle _A,
\end{eqnarray}
where besides the pure gauge link, the physical component
$\vec{A}_{phys}$ is used instead of $F_{\nu}^+$ as the gauge
invariant variable. The second moments of $\mathcal{P}_{g/A}$ and
$P_{g/A}$ relate to Poynting and the proper gluon momentum in
Eq.(12) and (39).

Our approach is also convenient in constructing the gauge invariant
polarized and transverse-momentum dependent PDFs with clear particle
number interpretation, and off-forward PDFs which can be measured to
infer the proper orbital angular momentums in Eq.(40). For example
the polarized gluon PDF can be defined gauge invariantly as
\begin{eqnarray}
P_{\Delta g/A}(\xi)= \int_{-\infty}^{\infty}  \frac{dx^-}{2\pi}
e^{-i\xi P^+x^-}\langle F^{+i}(0,x^-,0_\bot)
\mathcal{P}\exp\{ig\int_0^{x^-}dy^- A_{\rm
pure}^+(0,y^-,0_\bot)\}\epsilon_{ij+} A^j_{\rm phys}(0) \rangle _A,
\end{eqnarray}
with a first moment relating to the gauge invariant gluon spin in
Eq.(40).

\section{Summary and prospect}

Since the establishment of gauge invariance principle, we enjoy that
the total momentum, angular momentum and the Lorentz boosting
operators of a gauge system satisfy both the gauge invariance and
Poincar\'{e} algebra. However, we never have the separate momentum,
orbital angular momentum operators of the Fermion (electron in QED,
quark in QCD) and boson (photon in QED, gluon in QCD) part which
satisfy both the gauge invariance and the canonical momentum,
angular momentum Lie algebra. We have the electron, quark spin
operator but we never have the photon and gluon spin operators which
satisfy both the gauge invariance and spin Lie algebra. Even it had
been claimed in some textbooks that it is impossible to have a well
defined photon spin\cite{lab3}. The nucleon spin structure study
needs the gluon spin operator, but after about ten years effort in
searching for a gluon spin operator since the so-called proton spin
crisis, such an effort has almost been given up for the last ten
years. In this report we proposed a new set of quark (electron),
gluon (photon) momentum, orbital angular momentum and spin operators
which satisfy both the gauge invariance and the canonical momentum,
angular momentum Lie algebra.

To achieve this a key point is to separate the gauge field into pure
gauge and physical parts: the former one is unphysical and can be
gauged away as in Coulomb gauge, it is used to compensate the
induced unphysical gauge field due to the local gauge transformation
of the Fermion field to keep the gauge invariance; the physical part
is responsible for the physical coupling between Fermion and boson
field. It is physical and should be gauge invariant (covariant). We
provide a method to do this separation both for the Abelian $U(1)$
and the non-Abelian $SU(3)$ gauge field.

Our proposed momentum operators are different from the canonical
ones. The latter ones of the Fermion (electron in QED and quark in
QCD) are not gauge invariant and so do not represent observables
because they include the unphysical pure gauge field contribution.
The new ones subtract the unphysical pure gauge field contribution
and so they are gauge invariant and correspond the observables. The
Poynting vector used for the boson (photon in QED and gluon in QCD)
is not the proper momentum operator either because they do not
satisfy the momentum Lie algebra in the interacting field case.

We achieved to obtain a gauge invariant orbital angular momentum and
spin operators of the photon and gluon by means of the physical part
of the gauge field, Eq.(40), which provides the theoretical basis of
the widely used multipole radiation analysis, the photon spin and
orbital angular momentum used in quantum computation and
communication study, the gluon spin contribution in the nucleon spin
structure study.

In Coulomb gauge, the new momentum and orbital angular momentum
operators proposed in this paper coincide with the usually used
ones. This explains why the quantum mechanics calculations obtain
the right results. It is because usually these calculations are
performed in the Coulomb gauge. The multipole radiation calculation
is also performed in Coulomb gauge and so the results are correct
too.

The Poincar\'{e} algebra can not be fully maintained for the
momentum, angular momentum and Lorentz boosting operators of the
individual Fermion and boson part of an interacting gauge field
system. What is the meaning of these observables if they are not
Lorentz covariant? We have shown that the momentum and angular
momentum algebra can be maintained simultaneously with the gauge
invariance. How much part of the Poincar\'{e} algebra can be
maintained for the operators of the interacting Fermion and boson
separately, especially the Lorentz covariance can be maintained to
what extent are left for further study.

For a quantum gauge field system, in general one only requires the
matrix elements of the operator corresponding to a physical
observable between physical states to be gauge invariant. The strong
gauge invariance requirement might be a too strong one\cite{lab2}.
This should be studied further.

The new asymptotic limit of quark and gluon parton momentums of a
nucleon have been obtained\cite{lab7}, the immediate problem is the
asymptotic limit of the quark and gluon orbital angular momentums
and spins.

The gluon spin contribution to the nucleon spin is under measurement
in different labs. A lattice QCD calculation with the gauge
invariant gluon spin operator is called for.

How to relate the new PDFs to the measured cross sections in deep
inelastic scattering and virtual Compton scattering should be
examined.

In summary, the nucleon internal structure is better to be
reexamined based on the new quark, gluon momentum, orbital angular
momentum, spin operators and parton distribution functions and our
picture of the nucleon internal structure might be modified.

\acknowledgments{We thank Prof. X. D. Ji, K. F. Liu, J.W. Qiu, S. J.
Wang and Dr. J.P. Chen for stimulating discussions. This research is
supported by NSFC under Grant No. 90503011, 10875082, U.S. DOE under
Contract No. W-7405-ENG-36 and in part by the PKIP of CAS under
Grant No. KJCX2.YW.W10}

\end{document}